\documentclass[twocolumn,showpacs,amsfonts]{revtex4}
\usepackage{amsmath}
\usepackage{latexsym}
\usepackage{float}
\usepackage{amssymb}
\usepackage{graphicx}
\usepackage{textcomp}
\usepackage{hyperref}
\usepackage{graphicx}
\begin{document}
\def\be{\begin{equation}}
\def\ee{\end{equation}}     
\def\bfi{\begin{figure}}
\def\efi{\end{figure}}
\def\bea{\begin{eqnarray}}
\def\eea{\end{eqnarray}}

\newcommand{\bra}{\langle}
\newcommand{\ket}{\rangle}
\newcommand{\vk}{\vec{k}}
\newcommand{\rphi}{\widehat{\varphi}}
\newcommand{\iphi}{\widetilde{\varphi}}
\newcommand{\reta}{\widehat{\eta}}
\newcommand{\ieta}{\widetilde{\eta}}
\newcommand{\vphi}{\boldsymbol{\varphi}}
\newcommand{\eRk}{\emph{R}_{\vec{k},\eta}}
\newcommand{\eIk}{\emph{I}_{\vec{k},\eta}}

\title{Condensation of Fluctuations in and out of Equilibrium}

\author{Marco Zannetti$^{1,2}$, Federico Corberi$^{1}$ and Giuseppe Gonnella$^{3}$}
\affiliation {$^{1}$Dipartimento di Fisica ``E.R. Caianiello'', 
and CNISM, Unit\`a di Salerno,
Universit\`a  di Salerno, 
via Ponte don Melillo, 84084 Fisciano (SA), Italy. \\
$^{2}$Kavli Institute for Theoretical Physics China, CAS, Beijing 100190, China \\
$^{3}$Dipartimento di Fisica, Universit\`a di Bari and
INFN, Sezione di Bari, via Amendola 173, 70126 Bari, Italy}

\begin{abstract}

Condensation of fluctuations is an interesting phenomenon conceptually
distinct from condensation on average. One stricking feature is that, contrary to
what happens on average, condensation of fluctuations
may occurr even in the absence of interaction. 
The explanation emerges from the duality between large deviation events in the given system 
and typical events in a new and appropriately biased system.
This surprising phenomenon
is investigated in the context of the
Gaussian model, chosen as paradigmatical non interacting system, before and after
an istantaneous temperature quench. It is shown that 
the bias induces a mean-field-like effective interaction responsible of the condensation
on average. Phase diagrams, covering both the equilibrium and the off-equilibrium
regimes, are derived for observables representative of generic behaviors.
\pacs{05.70.Ln; 05.40.-a; 64.60.-i}
\end{abstract} 

\maketitle

\section{Introduction}
\label{I}

Condensation is a ubiquitous phenomenon in nature. It may take place
in equilibrium, off-equilibrium, in real space or in momentum space.
Starting from the most familiar condensation of
supersaturated vapour, there is a great abundance of
examples which includes, among others, the Bose-Einstein condensation (BEC)~\cite{Huang} and 
the related transition in mean-field systems, like the spherical~\cite{BK} or
the large-$N$ model~\cite{LN}. More recently there has
been much interest in condensation transitions arising out of equilibrium,
both in classical~\cite{ZPRreview} and quantum systems~\cite{Gambassi}.
In the non equilibrium context the phenomenology of condensation turns
out to be very rich with a variety of manifestations in fields as diverse as
economics, information theory, traffic models, granular materials, networks and
mass transport~\cite{ZPRreview,Marsili}. Much progress in the understanding of the basic features common to
most of these different instances of condensation has been achieved through the
study of driven diffusive systems and, in particular, of the zero-range process or
variants of it~\cite{ZPRreview}.

In this paper we shall focus on a yet another manifestation of condensation,
which is somewhat conceptually different. In the usual contexts mentioned above,
condensation is a phenomenon observed in the average behavior of the system. Instead,
we shall be concerned with condensation occurring in the {\it fluctuations}, namely with
condensation as a {\it rare} event~\cite{schutz,Kafri,FL,US,Evans}. The conceptual and substantial difference
is that condensation of fluctuations may occurr even in systems which cannot sustain
condensation on average, such as non interacting systems.
In order to emphasize this point, we shall work with the Gaussian model,
which is the paradigmatical non interacting system in the theory of phase transitions~\cite{Goldenfeld}.
Although the average properties
of this system are well known to be trivial, in and out of equilibrium, we shall
find that fluctuations of extensive quantities may condense.

Most of the work quoted above on condensation, both on average and in the fluctuations,
has been carried out in the context
of non equilibrium steady states, obtained by driving an externally generated
current into the system. Here, 
instead, we shall carry further the program initiated in Ref.~\cite{US}
of exploring fluctuations in the largely unknown area of the processes
wihout time translation invariance~\cite{ritort}. Specifically, we shall consider 
the ralaxation following the istantaneous quench from an initial temperature $T_I$ to a lower temperature $T_F$.
With such a choice, we can 
overview the entire evolution from the equilibrium behavior before the quench to the off equilibrium
relaxation after the quench. We shall see that, depending on the nature of the observable,
fluctuations may condense either in and out of equilibrium,
or just as an out of equilibrium phenomenon. 
We shall analyse in detail the mechanism of condensation and we shall derive
phase diagrams, extending into the time direction.
These diagrams show that during relaxation condensation is 
enhanced by the dynamics, if occurring also in equilibrium, or 
dynamically generated if absent in equilibrium.

The paper is organized as follows: In section~\ref{II} we set up the ensemble theory
apparatus needed in the rest of the paper.
The Gaussian model is introduced in section~\ref{III}. Section~\ref{IV} is the central section
of the paper, where the notions of condensation on average and condensation of the
fluctuations are discussed in general. The example of a macrovariable condensing both
in equilibrium and off equilibrium is treated in section~\ref{V}, while the example of
condensation as an out-of-equilibrium phenomenon is discussed in section~\ref{VI}. Concluding 
remarks are made in section~\ref{VII}.

\section{Ensembles}
\label{II}

The apparently puzzling feature of condensation appearing in the fluctuations
of a non interacting system
finds explanation in the framework of large deviation theory~\cite{Touchette}, 
through the mapping of rare fluctuations in the
given system (in our case the Gaussian model) into typical
events in a new system, obtained by the application of an appropriate bias. 
The key point, as we shall see, is that the bias 
produces an effective interaction, which is responsible of the condensation on average
in the biased system.
The basic idea amounts to an extension of ensemble
theory beyond the realm of equilibrium statistical mechanics,
according to a scheme which has been recently used in
a variety of different contexts, classical~\cite{derrida,nemoto,classical} and quantum~\cite{quantum}.

In order to give a general presentation of the method, let us consider a generic
probability distribution $P(\varphi,J)$, referred to as the {\it prior} and describing the state of
a system of volume V, with microstates consisting of sets of degrees of freedom $\varphi = [\varphi_i]$,
where $i$ is a generic label, and control parameters $J$.
In this paper $i$ is the position vector $\vec x$ in real space or the wave vector $\vec k$ in
Fourier space, and $J$ stands for temperature $T$ in equilibrium or for time $t$ off equilibrium.
The probability of a fluctuation $M$ of a random variable
${\cal M}(\varphi)$ is given by
\be 
P(M,J) = \int_{\Omega} d\varphi \, P(\varphi,J)\delta (M-{\cal M}(\varphi))
\label{gen.3}
\ee
where $\Omega$ is the phase space.
Introducing the integral representation of the $\delta$ function
$\delta(x) = \int_{\alpha-i\infty}^{\alpha + i\infty}\frac{dz}{2\pi i} \, e^{-zx}$ this becomes
\be
P(M,J) =  \int_{\alpha - i\infty}^{\alpha + i\infty} \frac{dz}{2\pi i} \, e^{-zM} K_{\cal M}(z,J)
\label{gen.5}
\ee
where
\be
K_{\cal M}(z,J) = \langle e^{z{\cal M}(\varphi)} \rangle
\label{gen.6}
\ee
is the moment generating function of ${\cal M}$ and
the brackets $\langle \cdot \rangle$ denote the average in the prior ensemble.
If the system is extended and ${\cal M}(\varphi)$ is an extensive macrovariable,
for large volume Eq.~(\ref{gen.5}) can be rewritten as
\be
P(M,J,V) =  \int_{\alpha - i\infty}^{\alpha + i\infty} \frac{dz}{2\pi i} \, e^{-V[zm + \lambda_{\cal M}(z,J)]}
\label{gen.6}
\ee
where $m$ is the density $M/V$ and
\be
-\lambda_{\cal M}(z,J) = \frac{1}{V}\ln K_{\cal M}(z,J,V)
\label{gen.7bis}
\ee 
is the volume independent scaled cumulant generating function. 
Carrying out the integration by the saddle point method,
the large deviation principle is obtained
\be
P(M,J,V) \sim e^{-VI_{\cal M}(m,J)}
\label{gen.8}
\ee
with the rate function 
\be
I_{\cal M}(m,J) = z^*m + \lambda_{\cal M}(z^*,J) 
\label{gen.9}
\ee
and where $z^*(m,J)$ is the solution, supposedly unique, of the saddle point equation
\be
\frac{\partial}{\partial z} \lambda_{\cal M}(z,J) = -m.
\label{gen.9bis}
\ee

From the above algebra follows the basic result of large deviation theory~\cite{Touchette} that
$I_{\cal M}(m,J)$ and $\lambda_{\cal M}(z,J)$ form a pair of Legendre transforms.
Therefore, regarding the latter 
quantity as the ``free energy'' of the new ensemble
\be
P(\varphi,z,J,V) = \frac{1}{K_{\cal M}(z,J,V)} P(\varphi,J,V) \, e^{z{\cal M}(\varphi)}
\label{Du.2}
\ee
obtained by imposing the exponential bias on the prior,
the rate function remains identified with the
``thermodynamic potential'' associated to yet another ensemble
\be
P(\varphi,M,J,V) = \frac{1}{P(M,J,V)} P(\varphi,J,V)\delta (M-{\cal M}(\varphi))
\label{gen.2}
\ee
in which the bias is implemented rigidly through
the phase space restriction $M={\cal M}(\varphi)$. 
To make contact with familiar ground, if the prior was the uniform ensemble $P(\varphi,V) = 1/|\Omega(V)|$
and ${\cal M}$ the energy of the system,
then $P(\varphi,z,V)$ and $P(\varphi,M,V)$ would be, respectively, the usual canonical ensemble at the
inverse temperature $\beta=-z$ and the microcanonical ensemble with energy $E=M$.

We stress that the above chain of relations holds in general, without limitations to equilibrium.
Therefore, the quantity $I_{\cal M}(m,J)$ 
plays two distinct roles~\cite{nemoto,Kafri,Evans}:
on the one hand it acts as the rate function regulating the occurrence of rare events in the
prior ensemble and on the other hand it is the thermodynamic potential accounting
for the average properties in the constrained ensemble $P(\varphi,M,J,V)$. In particular, if
the extra correlations due to the bias are responsible of singularities in the free energy,
amenable to a phase transition, the same singularities arise in the unbiased fluctuations. Consequently, 
the same phenomenon, in principle, could be observed following different experimental
protocols, either  by leaving the system unbiased and monitoring fluctuations
or, alternatively, by arranging the appropriate bias aimed to render typical the effect of interest.

\section{The Gaussian model}
\label{III}

In order to produce a concrete and simple realization of the above ideas, let us
consider the Gaussian model, which
describes a system of volume $V$, with a scalar order parameter
field $\varphi(\vec x)$ and the bilinear energy functional
\be
{\cal H}[\varphi] = \frac{1}{2} \int_V d \vec x \, [(\nabla \varphi)^2 + r \varphi^2 (\vec x)]
\label{GMD.1}
\ee
where $r$ is a non negative mass.
 The system is prepared in equilibrium at the temperature $T_I$. At the time $t=0$ is
istantaneously quenched to the lower temperature $T_F$. 
The dynamics, without conservation of the order parameter, are governed by the overdamped Langevin
equation~\cite{Goldenfeld,HH}
\be
\dot{\varphi}(\vec x, t) = \left [ \nabla^2 - r \right ] \varphi(\vec x, t)  + \eta(\vec x, t)
\label{Gin.4}
\ee
where $\eta(\vec x, t)$ is the white Gaussian noise generated by the cold reservoir, 
with zero average and correlator 
\be
\langle \eta(\vec x,t)\eta(\vec {x}^{\prime},t^{\prime}) \rangle = 
2T_F \delta(\vec x -\vec {x}^{\prime})   \delta(t-t^{\prime}).
\label{LG.2}
\ee
Due to linearity, the problem can be diagonalized by Fourier transformation.
Keeping in mind that the Fourier components
$\varphi_{\vec k} = \int_V d\vec x \, \varphi(\vec x) e^{i\vec k \cdot \vec x}$ are complex,
some care is needed in the identification of the independent variables.
Let us denote by ${\cal B}$ the set of all wave vectors with magnitude smaller than
an ultraviolet cutoff $\Lambda$, caused by the existence of a microscopic length scale
in the problem, like an underlying lattice spacing. 
Since the reality of $\varphi(\vec x)$ requires $\varphi_{-\vec k}=\varphi_{\vec k}^*$,
the independent variables are $\varphi_0$ 
and the set of pairs $ \{u_{\vec k} = \mathbb{R}e \varphi_{\vec k}, \; v_{\vec k} = \mathbb{I}m \varphi_{\vec k} \}$
with $\vec k \in {\cal B}_+$, where ${\cal B}_+$ is a half of ${\cal B}$.
More precisely, if ${\cal B}_-$ is the set obtained by reversing all the wave vectors in ${\cal B}_+$,
then ${\cal B}_+$ is such that ${\cal B}_+  \cap {\cal B}_- = \emptyset$ and 
${\cal B}_+  \cup {\cal B}_- = {\cal B}-\{ \vec 0 \}$.
However, rather than working with ${\cal B}_+$, it is more convenient to
let $\vec k$ to vary over the whole of ${\cal B}$ by taking as independent real variables
\be
x_{\vec k}  = \left \{ \begin{array}{ll}
         \varphi_0 ,\;\; $for$ \;\; \vec k = 0  ,\\
         u_{\vec k} ,\;\; $for$ \;\; \vec k \in {\cal B}_+, \\
         v_{\vec k} ,\;\; $for$ \;\; \vec k \in {\cal B}_-.
        \end{array}
        \right .
        \label{frrr.3}
        \ee
With this convention, from Eq.~(\ref{Gin.4}) we get
the equations of motion for a set of independent Brownian oscillators
\be
\dot{x}_{\vec k}(t) = -\omega_k x_{\vec k}(t) + \zeta_{\vec k}(t)
\label{FOUR.2}
\ee 
with the dispersion relation $\omega_k = (k^2 + r)$. The noise correlator is given by
\be
\langle \zeta_{\vec k}(t) \zeta_{\vec{k}^{\prime}}(t^{\prime}) \rangle  = 2T_{F,k} V \delta_{\vec k,-\vec{k}^{\prime}}\delta(t-t^{\prime})
\label{FOUR.8}
\ee
where
\be
T_{F,k} = \frac{T_F}{2\theta_k}
\label{finT.1}
\ee
and $\theta_k$ is the Heaviside step function with $\theta_0=1/2$.
The energy functional~(\ref{GMD.1}) then takes the separable form
\be
{\cal H}(\mathbf{x}) = \sum_{\vec k}{\cal H}_{\vec k}(x_{\vec k})
 \label{GMD.2bis}
\ee
with
\be
{\cal H}_{\vec k}(x_{\vec k}) = \frac{1}{V}\theta_k \omega_k x^2_{\vec k}
\label{FOUR.9}
\ee
and where $\mathbf{x}$ stands for the whole set $\{x_{\vec k}\}$.

\begin{figure}[h]

	\vspace{1cm}

    \centering
   \rotatebox{0}{\resizebox{.35\textwidth}{!}{\includegraphics{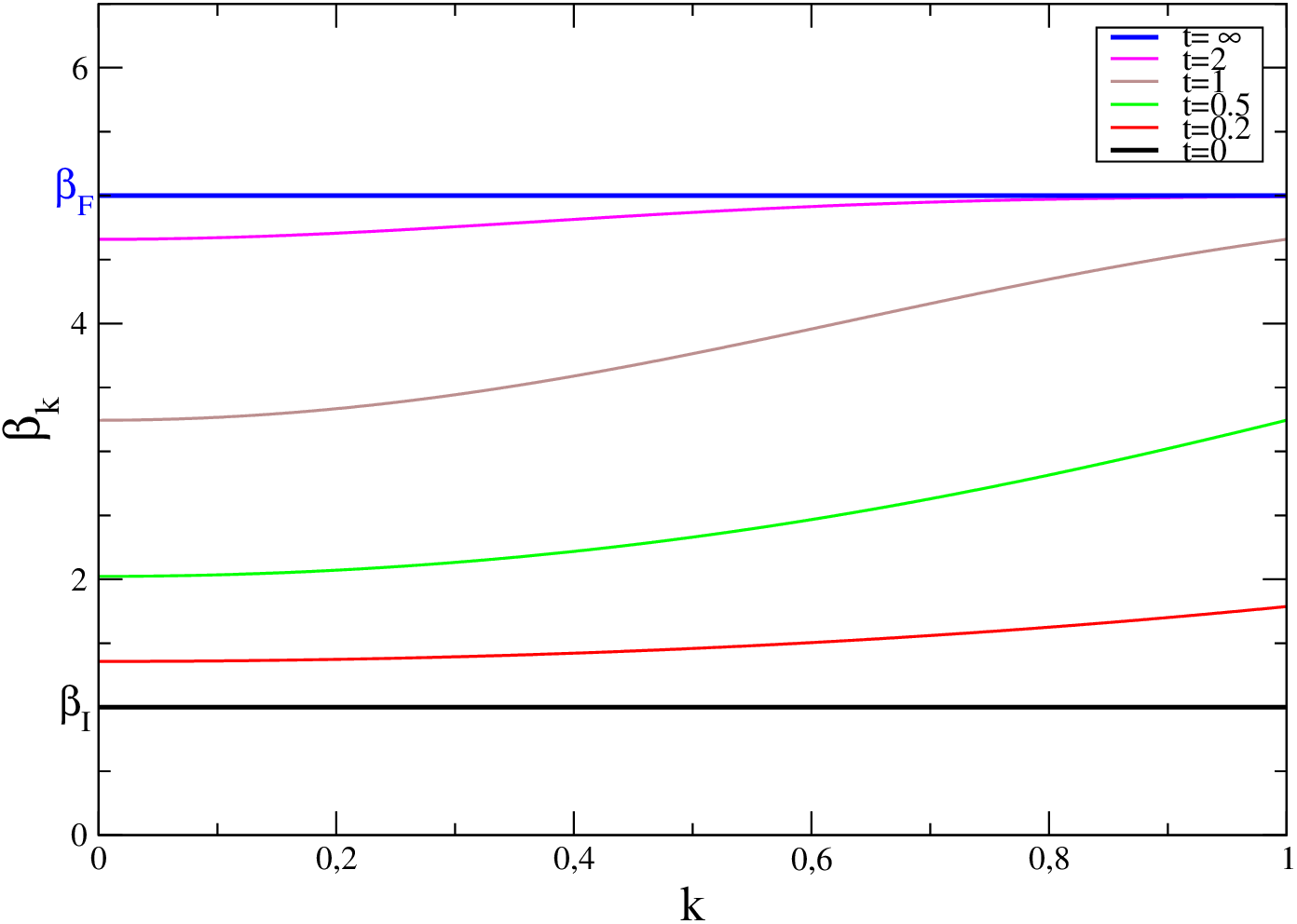}}} \vskip 0.7cm
   \rotatebox{0}{\resizebox{.35\textwidth}{!}{\includegraphics{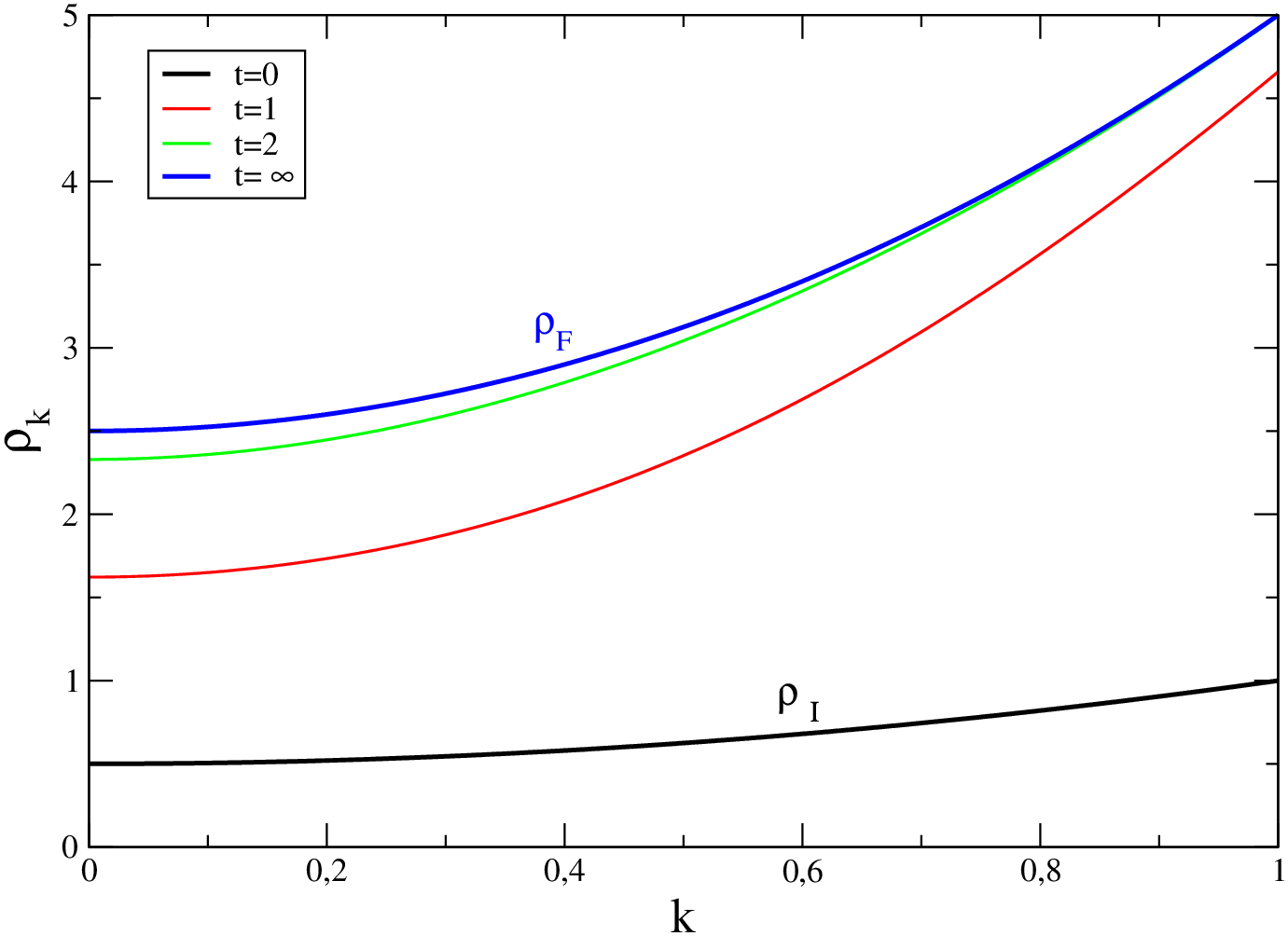}}}
   \caption{Spectra of inverse effective temperatures (top) and of $\rho_k$ for
the order parameter sample variance (bottom), with $r=1,T_I=1,T_F=0.2$.}
\label{effT}
\end{figure}

Due to mode independence, the state of the system is factorized at all times
$P(\mathbf{x},t) = \prod_{\vec k} P_{\vec k}(x_{\vec k},t)$, with
the single-mode contributions given by
\be
P_{\vec k}(x_{\vec k},t)  =  Z^{-1}_{\vec k}(t) e^{-\beta_k(t) {\cal H}_{\vec k}(x_{\vec k})}
\label{Indv.1}
\ee
\be
Z_{\vec k}(t)=\sqrt{\frac{\pi V}{\beta_k(t)\theta_k \omega_k}}
\label{Indv.2}
\ee
where $\beta^{-1}_k(t)$ is the effective temperature of the modes with wave vector magnitude $k$, 
defined from the average energy per degree of freedom~\cite{Ciliberto}   
\be
\beta^{-1}_k(t) =   2\langle {\cal H}_{\vec k}(t) \rangle = \frac{2}{V} \theta_k \omega_k
\langle x_{\vec k}^2(t) \rangle
\label{Indv.2bis}
\ee
which yields
\be
\beta^{-1}_k(t) = (T_I-T_F) e^{-2\omega_k t} + T_F.
\label{1}
\ee  
In this paper we shall take $k_B=1$ for the Boltzmann constant. 
As illustrated in the top panel of Fig.\ref{effT}, initially the spectrum of effective temperatures
is flat with $\beta_k(t=0) = \beta_I$,
which is the statement of energy equipartiton.
Then, as the system relaxes, the temperatures of the different modes
acquire a $k$-dependence, signaling the breaking of
equipartion and departure from equilibrium. Eventually, convergence to the same final value $\beta_F$
takes place, as the system equilibrates and equipartition is restored. 
The probability distribution $P(\mathbf{x},t)$ will be taken as the prior in the following.

\section{Fluctuations of a macrovariable}
\label{IV}

Let us now consider a quadratic and separable macrovariable 
${\cal M}(\mathbf{x}) = \sum_{\vec k}{\cal M}_{\vec k}(x_{\vec k})$, with
${\cal M}_{\vec k}(x_{\vec k}) = \frac{1}{V}\theta_k \mu_k x_{\vec k}^2$, whose
coefficients $\mu_k$ are to be specified. According to the scheme of section~\ref{II},
all the information on the fluctuations of ${\cal M}(\mathbf{x})$ at the
generic time $t$ is contained in the rate function~(\ref{gen.9}), with $J=t$.
The computation of this quantity requires the preliminary computation of
the moment generating function. From the factorization of the prior and the
separability of ${\cal M}$ follows 
\be
K_{{\cal M}}(z,t) =   \prod_{\vec k}K_{{\cal M},\vec k}(z,t)
\label{EnHe.1}
\ee
with the single-mode factors given by
\begin{eqnarray}
K_{{\cal M},\vec k}(z,t) & = &  \int_{-\infty}^{\infty} dx_{\vec k} \, P_{\vec k}(x_{\vec k},t)
e^{z{\cal M}_{\vec k}(x_{\vec k})} \nonumber \\ 
& = &   \frac{1}{\sqrt{1 - \rho^{-1}_k(t)z}}
\label{EnHe.5}
\end{eqnarray}
where 
\be
\rho_k = \beta_k\omega_k/\mu_k  = \frac{1}{2}\langle {\cal M}_{\vec k} \rangle^{-1}.
\label{ro.1}
\ee
Inserting this result into Eq.~(\ref{gen.7bis}), the saddle point equation~(\ref{gen.9bis})
can be written as
\be
m = \widetilde{F}_{{\cal M}}(z,t,V) 
\label{MDF.4}
\ee
where  the function in the right hand side is given by
\be
\widetilde{F}_{{\cal M}}(z,t,V) = \frac{1}{V} \sum_{\vec k} \langle {\cal M}_{\vec k} \rangle_z
\label{MDF.6}
\ee
and
\be
\langle {\cal M}_{\vec k} \rangle_z = \frac{1}{2[\rho_k(t) - z]}
\label{MDF.6bis}
\ee
is the average per mode in the biased ensemble~(\ref{Du.2}). Recalling the definition~(\ref{ro.1}) of $\rho_k$, the above
equation can be rewritten as 
\be
\langle {\cal M}_{\vec k} \rangle_z = \frac{1}{\langle {\cal M}_{\vec k}(t) \rangle^{-1} -  2z}
\label{Dyson.5}
\ee
in which the biased and the prior averages enter in the same formal
relationship as the dressed and the bare average in a Dyson-Schwinger-type equation~\cite{Amit,Zin}, 
with $2z$ playing the role of the tadpole self-energy. Now, since truncating the self-energy
skeleton expansion to the tadpole contribution amounts to a self-consistent mean-field approximation, as in the
large $N$ limit of an $O(N)$ model~\cite{Ma,Zin}, we have that biased expectations can
be viewed as arising from the mean-field approximation on an underlying interacting theory,
whose free limit is given by the prior expectations. This turns out to be essential for
the distinction between condensation as a typical phenomenon or as a rare fluctuation.

\begin{figure}[h]

	\vspace{1cm}

    \centering
   \rotatebox{0}{\resizebox{.35\textwidth}{!}{\includegraphics{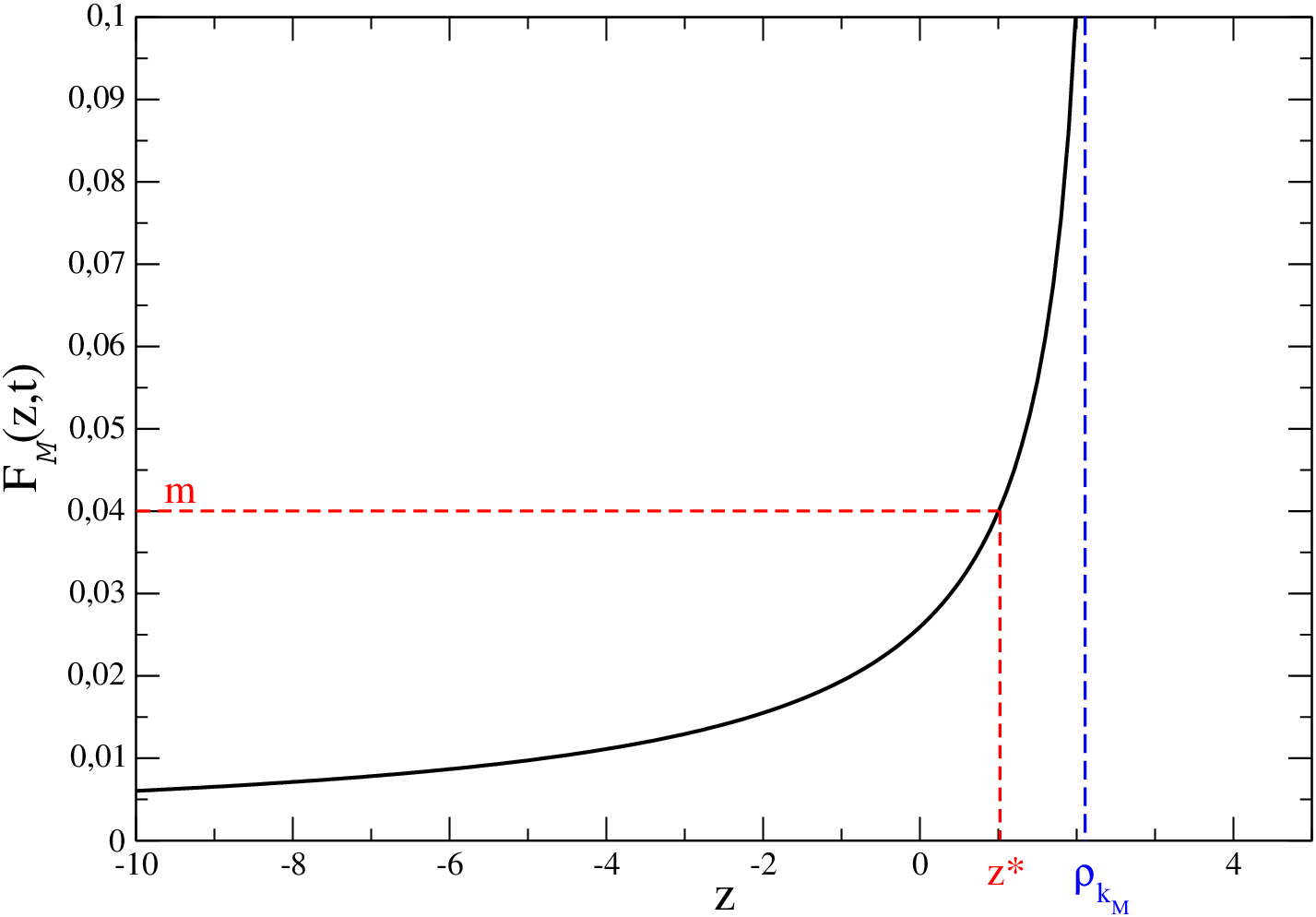}}} \vskip 0.7cm
   \rotatebox{0}{\resizebox{.35\textwidth}{!}{\includegraphics{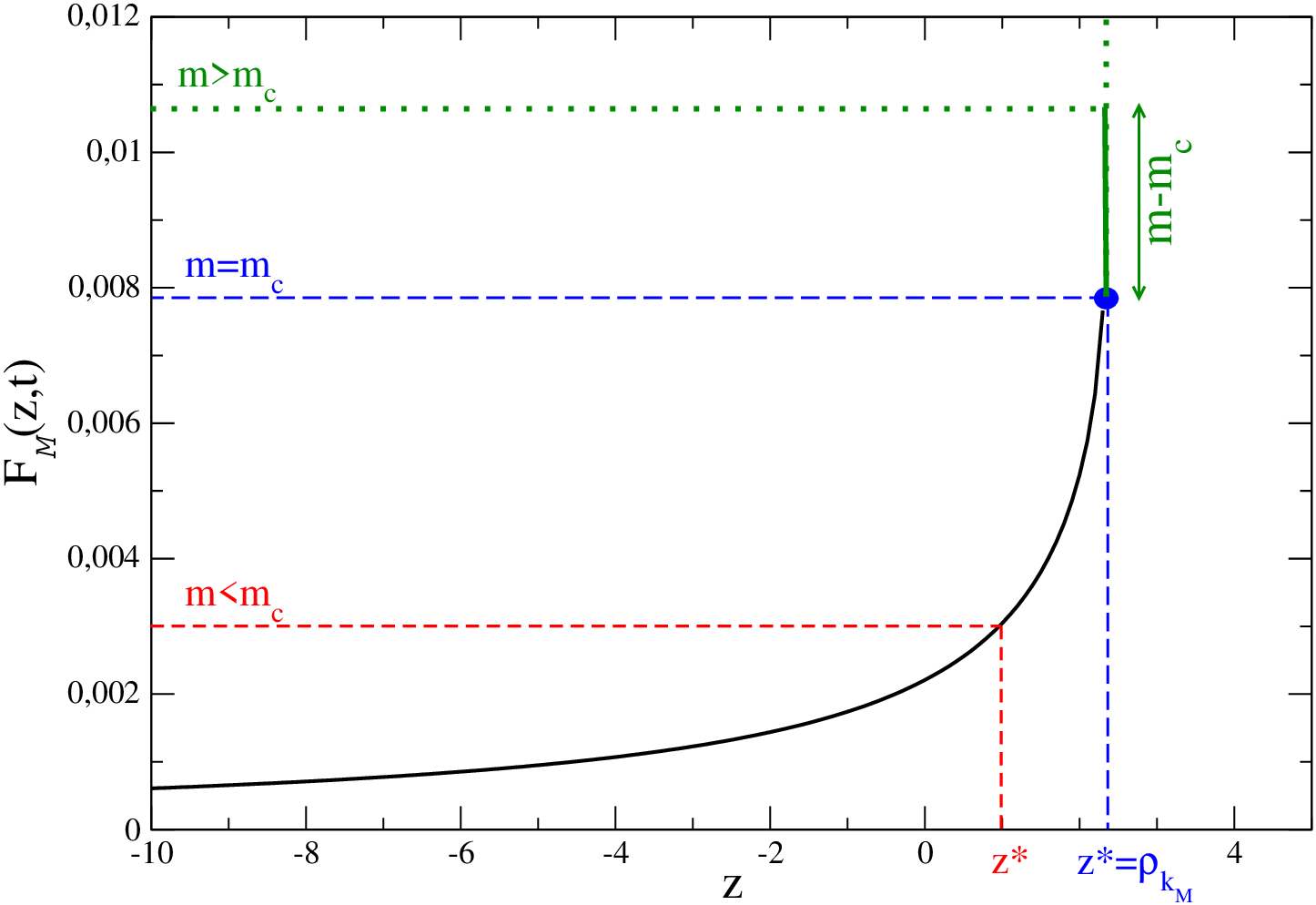}}}
   \caption{Typical behavior of $F_{\cal M}(z,t)$, obtained with $\mu_k=2, t=2, r=1,T_I=1,T_F=0.2$.
Top panel: graphical solution of Eq.~(\ref{F.2}) with $d =1$.
Bottom panel: graphical solution of Eq.~(\ref{4}), with $d =3$ and for $m < m_C$, $m = m_C$ 
and $m > m_C$.}
\label{Fofz}
\end{figure}

Transforming the sum in Eq.~(\ref{MDF.6}) into an integral, the saddle point equation~(\ref{MDF.4}) 
can be rewritten as
\be 
m=F_{\cal M}(z,t)
\label{F.2}
\ee 
with
\be
F_{\cal M}(z,t) = \frac {\Upsilon_d}{2} \int_0^{\Lambda} \frac{dk}{(2 \pi)^d} \, \frac{k^{d-1}}{\rho_k(t) - z}
\label{F.1}
\ee
where $d$ is the space dimensionality, $\Upsilon_d=2\pi^{d/2}/\Gamma (d/2)$ the $d$-dimensional solid angle and
$\Gamma$ the Euler gamma function. 
The formal solution is given by
\be
z^*(m,t) = F_{{\cal M}}^{-1}(m,t)
\label{sol.1}
\ee
where $F_{{\cal M}}^{-1}$ 
is the inverse, with respect to $z$, of the function defined by Eq.~(\ref{F.1}).
The existence of this solution
depends on the domain of definition of  $F_{\cal M}^{-1}$.
If we assume ${\cal M}$ to be positive, $F_{\cal M}^{-1}$ is defined for
$z \leq \rho_{k_M}$, where $k_M$ is the wave vector which minimizes $\rho_k$, and 
\be
F_{\cal M}(z,t) \leq m_C(t)
\label{sol.2}
\ee
with
\be
m_C(t) = F_{\cal M}(z=\rho_{k_M},t). 
\label{uppb.1}
\ee
The issue is whether this upper bound
is finite or infinite. 
In this paper, for simplicity, we shall limit the discussion to cases with
$k_M = 0$. Then, if $[\rho_k(t) - \rho_0(t)]$ vanishes with $k$
like $k^{\alpha}$, for $d \leq \alpha$ the singularity is not integrable,
$m_C(t)$ diverges and the solution~(\ref{sol.1}) exists for any $m \geq 0$. This is
shown in the top panel of Fig.~\ref{Fofz}. Instead, if
$d > \alpha$, the singularity is integrable, $m_C(t)$ is finite and
the solution~(\ref{sol.1}) 
exists only for $m \leq m_C(t)$ (bottom panel of Fig.~\ref{Fofz}). In order to find the solution for 
$m > m_C(t)$ one must proceed as in the standard treatment of BEC~\cite{Huang}, separating the $k=0$
term from the sum and rewriting Eq.~(\ref{F.2}) as 
\be
m = \frac{1}{V} \langle {\cal M}_0 \rangle_{z^*} + F_{\cal M}(z^*,t).
\label{4}
\ee
Then, $m_C(t)$ defines a critical line on the
$(t,m)$ plane separating the normal phase (below) from the condensed phase (above).
Below, the first term in the right hand side of
Eq.~(\ref{4}) is ${\cal O}(1/V)$ and negligible, while above (see Fig.~\ref{Fofz}) takes
the finite value $[m - m_C(t)]$, due to the ``sticking''~\cite{Huang,BK}
of $z^*$ to the $m$-independent value $z^*=\rho_0(t)$. Summarising,
\be
z^*(m,t)  = \left \{ \begin{array}{ll}
        F_{\cal M}^{-1}(m,t) ,\;\; $for$ \;\; m \leq m_C(t),\\
        \rho_0(t) ,\;\; $for$ \;\;  m > m_C(t),
        \end{array}
        \right .
        \label{ODG.6}
        \ee
as it is illustrated in the bottom panel of Fig.~\ref{Fofz}.
What we have derived, so far, is condensation on average in the framework of the biased ensemble.
That is, the transition from microscopic to macroscopic 
of the expectation  $\langle {\cal M}_0 \rangle_{z^*}$,
analogous to BEC for the zero momentum occupation number. We emphasize, for future reference, that
the occurrence of the transition requires i) that the intensive parameter $\rho$ conjugate to
${\cal M}$ does depend on $k$, i.e. that there exists a spectrum of parameters $\rho_k$ and ii)
that the spectrum vanishes with $k$ as $k^{\alpha}$ with $\alpha < d$.

\begin{figure}[h]

	\vspace{1cm}

    \centering
   \rotatebox{0}{\resizebox{.35\textwidth}{!}{\includegraphics{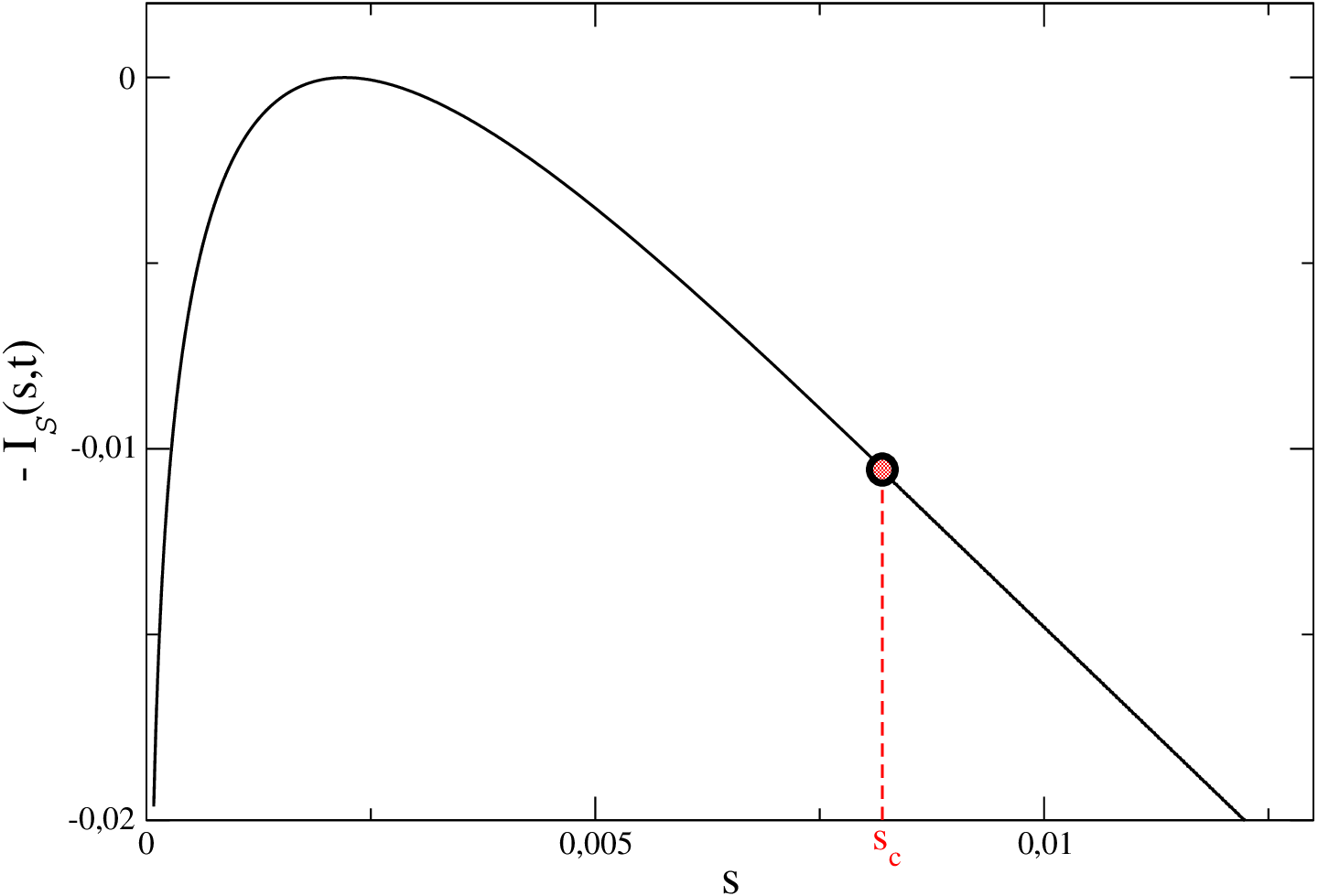}}} \vskip 0.7cm
   \caption{Rate function $I_{\cal S}(s,t)$ for the sample variance discussed in section~\ref{V}, 
$s_C$ denotes the critical threshold. Parameters $\mu_k=2, t=2, r=1,T_I=1,T_F=0.2$, $d=3$.}
\label{Ratefunction}
\end{figure}

In order to see the dual image of this transition
in the fluctuations occurring in the prior
ensemble~\cite{heat}, we must take a look at the rate function. Taking into account the definition~(\ref{gen.9})
and the above result for $z^*(m,t)$, we have
\be
I_{\cal M}(m,t)  = \left \{ \begin{array}{ll}
        z^*(m,t)m + \lambda_{\cal M}(z^*(m,t),t), \;\; $for$ \;\; m \leq m_C(t),\\
        \rho_0(t)(m-m_C) + I_{\cal M}(m_C,t), \;\; $for$ \;\;  m > m_C(t),
        \end{array}
        \right .
        \label{I.1}
        \ee
whose typical behavior is displayed in Fig.~\ref{Ratefunction}, obtained for the sample variance
discussed below in section~\ref{V}.
Thus, the probability of a fluctuation with $m > m_C(t)$ is given by
\be
P(M,t) \sim e^{-V\rho_0(t)(m-m_C)}e^{-VI_{\cal M}(m_C,t)}.
\label{I.1}
\ee
On the other hand, the fluctuations probability can also be written as
\be
P(M,t) = \int \prod_{\vec k} dM_{\vec k} \,
P(\{M_{\vec k}\},t) \delta (M- \sum_{\vec k} M_{\vec k})
\label{condfluct.1}
\ee
where $\{M_{\vec k}\}$ is a configuration of the values taken by the single-mode observables
${\cal M}_{\vec k}$. 
The statement is simply that, once $M$ has been fixed,
the allowed microscopic events $\{M_{\vec k}\}$ are those on the hypersurface defined by the constraint
$M = \sum_{\vec k} M_{\vec k}$ and that the probability $P(M,t)$ is obtained by summing
over the shell.
The probability of one such configuration is given by
\be
P(\{M_{\vec k}\},t) =  \prod_{\vec k} P_{\vec k}(M_{\vec k},t)
\label{condfluct.2}
\ee
where $P_{\vec k}(M_{\vec k},t)$, appearing in the right hand side, using
Eqs.~(\ref{gen.5}) and~(\ref{EnHe.5},) is given by
\be 
P_{\vec k}(M_{\vec k},t) = \frac{e^{-\rho_kM_{\vec k}}}{\sqrt{ \pi \rho^{-1}_k M_{\vec k}}} \theta(\rho^{-1}_k M_{\vec k})
\label{hr.10}
\ee 
and $\theta$ is, again, the Heaviside step function.
Now, inserting this result into Eq.~(\ref{condfluct.1}) and comparing with Eq.~(\ref{I.1}),
we obtain
\begin{eqnarray}
P(M,t) & = & \int dM_{0} P_0(M_{0},t)\delta (M_0- (M-M_C)) \nonumber \\
& \times & \int \prod_{\vec k \neq 0} dM_{\vec k} \,
P_{\vec k}(M_{\vec k},t) \delta (M_C- \sum_{\vec k \neq 0} M_{\vec k}) \nonumber
\label{condfluct.10}
\end{eqnarray}
which means that, for $m > m_C(t)$, the probability of the configurations $\{ M_{\vec k}\}$
is concentrated on the subset of the shell singled out by 
the additional condition $M_0 = M - M_C$.
This is condensation of fluctuations, in the sense that a fluctuation above the threshold $M_C$
can occurr only if the macroscopic fraction $M-M_C$ of it is contributed by the zero mode.
As anticipated in section~\ref{I}, the remarkable feature of this transition is that it
takes place in a non interacting system, like the Gaussian model, in which
no transition on average can take place, in and out of equilibrium. The explanation is in Eq.~(\ref{Dyson.5}),
which shows how the bias generates the interaction sustaining the transition, and the bias is
generated once the size of the fluctuation has been fixed.

As an illustration, in the next sections we shall analyse two specific cases. 
In the first one condensation occurs both in equilibrium and out of
equilibrium, while in the second one it occurs exclusively as an out of equilibrium phenomenon.

\section{Order parameter sample variance}
\label{V}

\begin{figure}[h]

	\vspace{1cm}

    \centering
   \rotatebox{0}{\resizebox{.4\textwidth}{!}{\includegraphics{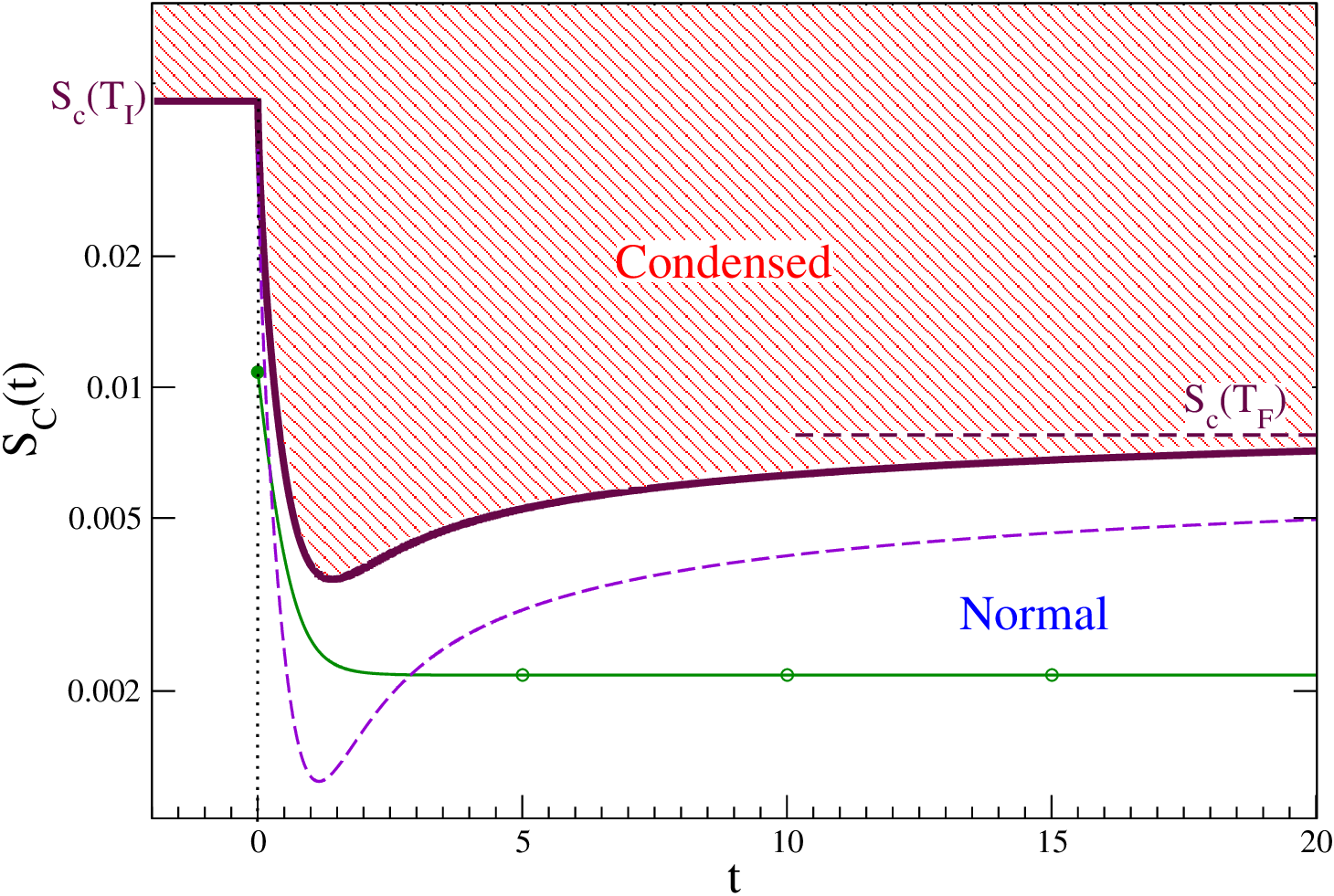}}}\vskip 0.7cm
   \rotatebox{0}{\resizebox{.4\textwidth}{!}{\includegraphics{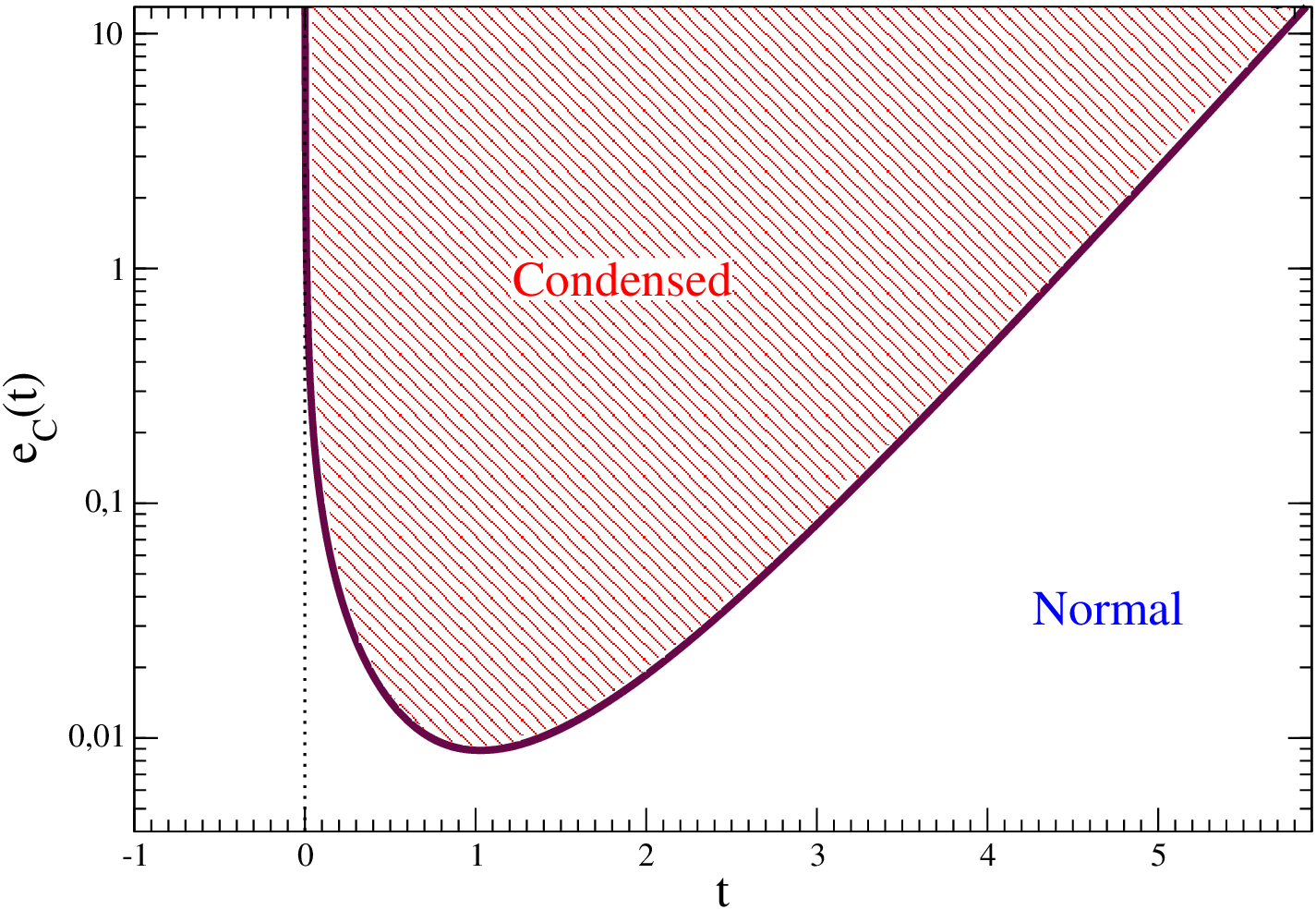}}}
   \caption{Top panel: Phase diagram of order parameter sample variance.
The upper horizontal dashed line corresponds to $s_C(T_F)$. The green line is the plot of
$\langle s(t) \rangle$. The lower dashed line is the difference $[s_C(t) - \langle s(t) \rangle]$.
Bottom panel: Energy phase diagram. In both cases: $r=1,T_I=1,T_F=0.2, d=3$.}
\label{ph_diag_S}
\end{figure}

Let us consider the sample variance
\be
{\cal S}[\varphi] = \int_V d\vec x \, \varphi^2(\vec x) = 
\frac{1}{V}\sum_{\vec k} \theta_k x_{\vec k}^2
\label{sph.1}
\ee
as the fluctuating macrovariable.
This corresponds to $\mu_k=2$, which is independent of $k$ and yields $\rho_k(t) = \beta_k\omega_k/2$.
From the small $k$ behavior $[\rho_k(t) - \rho_0(t)] \sim k^2$
follows $\alpha = 2$ for all times, including the initial and the final equilibrium states 
(bottom panel of Fig.~\ref{effT}).
Therefore, denoting by $s$ the density $S/V$, the critical value $s_C(t)$
is finite for $d > 2$ at all times. The critical line for $d=3$ 
is displayed in the top panel of Fig.~\ref{ph_diag_S}.
In order to understand this phase diagram, 
one should keep in mind that fixing the value of $s$ amounts to implement
a {\it spherical} constraint {\it \`a la} Berlin and Kac~\cite{BK}. 
Let us first consider equilibrium, in the time region $t \leq 0$ preceding
the quench. Here, the critical line 
is horizontal and corresponds to the critical threshold $s_C(T_I)$ of the spherical
model at the temperature $T_I$~\cite{driveBK}.
Then, according to the dual point of view expounded above, the two
alternative readings of the equilibrium transition are that condensation
can be observed either as the usual transition of the spherical model
or as a rare event in the Gaussian model, where the sample variance is free to fluctuate.

Consider, next, the relaxation regime after the quench, for $t > 0$.
As it is evident form Fig.~\ref{ph_diag_S}, 
there are two time regimes separated by the minimum of the critical line,
about the characteristic time $\tau \sim r^{-1}$, which is 
the relaxation time of the slowest mode. In the first regime $(0 < t < \tau)$ the system
is strongly off equilibrium and the threshold $s_C(t)$ drops abruptly.
In the second regime $(t > \tau)$ the system gradually equilibrates
to the final temperature and $s_C(t)$ saturates slowly toward the final equilibrium 
value $s_C(T_F) < s_C(T_I)$.
A few observations are in order: 
i) The plot of the unbiased average $\langle s(t) \rangle$ lies
below the critical line, showing that condensation
of fluctuations is always a rare event. 
However, the plot of $[s_C(t) - \langle s(t) \rangle]$ shows
that the {\it rarity} of the condensation event varies with time and that the most favourable
time window for condensation is around $\tau$, where the difference is minimized.
Hence, condensation of the fluctuations is enhanced by the off equilibrium dynamics.
ii) The nonmonotonicity of the critical line is a remarkable dynamical feature, leading
to a re-entrance phenomenon. Namely, when the transition is driven by $t$, 
and $s$ is kept fixed to a value in between $s_C(T_F)$ and $s_C(T_I)$, a fluctuation of this size at first
is normal and  then condenses,  while for $s$ in between the minimum of the critical line and $s_C(T_F)$, 
the fluctuation undergoes a second and reverse transition becoming  normal again at late times.  iii)
The dynamical condensation here analysed is not related
to the properties of the dynamical spherical model~\cite{Godreche}, 
which requires the spherical
constraint to be imposed pathwise, namely at all times after the quench. 
Here, instead, the evolution is unconstrained and the spherical constraint
is imposed only at the observation time $t$. Therefore, while in equilibrium the two experimental protocols,
fluctuations monitoring vs bias implementation, are in principle both realizable, the latter one requiring
an istantaneous bias is hardly realizable off equilibrium.

\section{Energy} 
\label{VI}

As a second example, let us consider the energy~(\ref{GMD.2bis}) as the fluctuating macrovariable.
This is representative of a different
class of observables, whose fluctuations behave normally in equilibrium and
undergo a condensation transition  
as an out of equilibrium phenomenon. This is due to
$\mu_k = \omega_k$, from which follows $\rho_k=\beta_k$. Therefore, the
$k$ dependence of $\rho_k$ disappears in equilibrium (Fig. \ref{effT}) shifting to infinity
the critical threshold. More in detail, denoting by $e$ the energy density, 
the critical line is given by
\be 
e_C(t) = \int_0^{\Lambda} \frac{dk}{4\pi^2} \, \frac{k^2}{\beta_k(t) - \beta_0(t)}.
\label{criten.1}
\ee
The corresponding phase diagram, in the bottom panel of Fig. \ref{ph_diag_S}, is qualitatively different from the 
one in the top panel for the absence of the phase
transition in equilibrium.  This is due to the fact that, in equilibrium,
the denominator $(\beta_k - \beta_0)$ under the integral
vanishes identically for all $k$. This implies $\alpha \rightarrow \infty$ and the divergence of both $e_C(T_I)$ and
of $e_C(T_F)$ for any space dimensionality $d$. However, as soon as the system is put off equilibrium,
equipartion is broken
and the spectrum of inverse effective temperatures develops a minimum at 
$k=0$ (Fig. \ref{effT}). Then, the integral becomes convergent for $d > 2$. Consequently, $e_C(t)$ 
drops down from infinity to a minimum around
$\tau$ and, then, rises again toward infinity as the system reaches the final equilibrium state.
The nonmonotonic shape of the critical line implies, also in this case, re-entrance
of the $t$-driven transition for all fluctuations above the minimum of the critical line.

\section{Conclusions}
\label{VII}

In summary, we have analysed the behavior of fluctuations of macrovariables in
the Gaussian model, both in equilibrium and in the off equilibrium relaxation
following a sudden temperature quench. 
 For a certain class of bilinear variables
there is condensation
in the behavior of large deviations, in the sense that the $k=0$ mode contributes
a macroscopic amount to the fluctuations. The transition in the fluctuations is dual
to an ordinary transition, sustained by an effective mean-field interaction, in the constrained 
or biased system.
Differently from previous work on condensation, 
we have considered  equilibrium follwed by relaxation through a non
stationary process, in which the time evolution plays an essential role.
Also, essential is the {k}-space structure of macrovariables and the dispersion relation
in the prior model, which is a feature not present in models with identically distributed variables \cite{Evans}.
In this respect, it is particularly interesting the case of energy fluctuations
as an instance in which the $k$ dependence of the conjugate intensive parameters $\rho$
is dynamically generated and, with it, also the occurrence of condensation.
Finally, duality is a general property, not limited to the case of
a non interacting prior. Future work will be devoted
to the investigation of fluctuations singularities in the case of interacting systems.

\noindent {\it e-mail addresses} - mrc.zannetti@gmail.com, corberi@sa.infn.it, gonnella@ba.infn.it


\begin{thebibliography}{99}
 
\bibitem{Huang}
K. Huang, {\it Statistical Mechanics}, John Wiley and Sons, New York 1967

\bibitem{BK}
T. H. Berlin and M. Kac, Phys. Rev. {\bf 86}, 821 (1952)

\bibitem{LN}
For the condensation transition when the spherical constraint is imposed in the mean
via the large $N$ limit, see
C. Castellano, F. Corberi, and M. Zannetti, Phys. Rev. E {\bf 56}, 4973 (1997)

\bibitem{ZPRreview}
For a review on condensation in driven diffusive systems see
M. R. Evans and T. Hanney, J. Phys. A: Math. Gen. {\bf 38}, R195 (2005)
and references quoted therein; M. R. Evans and B. Waclaw, J. Phys. A: Math. Theor. {\bf 47}, 095001 (2014).

\bibitem{Gambassi}
A. Gambassi and A. Silva, Phys. Rev. Lett. {\bf 109}, 250602 (2012)

\bibitem{Marsili}
M. Filiasi, G.Livan, M. Marsili. M. Peressi, E. Vesselli and E. Zarinelli, arXiv:1201.2817v1; 
M. Filiasi, E. Zarinelli, E. Vesselli and M. Marsili, arXiv:1309.7795v1;
L. Ferretti, M. Mamino and G. Bianconi, arXiv:1310.3852v1

\bibitem{schutz}
R.J. Harris, A. R\'akos, and G.M. Schuetz, J. Stat. Mech. P08003 (2005)

\bibitem{Kafri}
N. Merhav and Y. Kafri, J. Stat. Mech. P02011 (2010)

\bibitem{US}
F. Corberi, G. Gonnella, A. Piscitelli and M. Zannetti, J. Phys. A: Math. Theor. {\bf 46}, 042001 (2013)

\bibitem{FL}
F.Corberi and L.F.Cugliandolo, J. Stat. Mech. P11019 (2012)

\bibitem{Evans}
J. Szavits-Nossan, M. R. Evans and S. N. Majumdar, Phys. Rev. Lett. {\bf 112}, 020602 (2014).

\bibitem{Goldenfeld}
N. Goldenfeld, {\it Lectures on Phase Transitions and the Renormalization Group}, Addison-Wesley Publishing Co.,
Reading, Mass. 1992;
P. M. Chaikin and T. C. Lubenski, {\it Principles of Condensed Matter Theory}, Cambridge University Press 1995

\bibitem{ritort}
A. Crisanti and F. Ritort, Europh. Lett. {\bf  66}, 253 (2004).

\bibitem{Touchette}
H. Touchette, Phys. Rep. {\bf 478}, 1 (2009)

\bibitem{derrida}
F. Ritort, J. Stat. Mech.: Theory and Experiment, P10016 (2004);
B. Derrida, J. Stat. Mech. P07023 (2007); C. Jardina, J. Kurchan and L. Peliti, Phys. Rev. Lett. {\bf 96},
120603 (2006); C. Jardina, J. Kurchan, V. Lecomte and J. Tailleur, J. Stat. Phys. {\bf 145}, 787 (2011).

\bibitem{nemoto}
T. Nemoto and S. Sasa, Phys. Rev. E {\bf 84}, 061113 (2011) and arXiv:1309.7200v2.

\bibitem{classical}
R. Jack and P. Sollich, Progr. Theor. Phys. Supp. {\bf 184}, 304 (2010); E. S. Loscar, A. S. J. S. May and
J. Garrhan, J. Stat. Mech.: Theory and Experiment, (2011) P12011; 
R. Chetrite and H. Touchette, Phys. Rev. Lett. {\bf 111}, 120601 (2013); A. A. Budini, R. M. Turner and J. P. Garrahan,
arXiv:1311.1031v1

\bibitem{quantum}
J. M. Hickey, S. Genway and J. P. Garrahan, arXiv:1309:1673v1; D. Manzano and P. I. Hurtado, arXiv:1310.7370v1

\bibitem{HH}
P. C. Hoenberg and B. I. Halperin, Rev. Mod. Phys. {\bf 49}, 435 (1977).

\bibitem{Ciliberto}
J. R. Gomez-Solano, A. Petrosyan and S. Ciliberto, Phys. Rev. Lett. {\bf 106}, 200602 (2011)

\bibitem{Amit}
D. J. Amit and M. Zannetti, J. Stat. Phys. {\bf 7}, 31 (1973).

\bibitem{Zin}
J. Zinn-Justin, {\it Quantum Field Theory and Critical Phenomena}, Chpt. 30, 4th Edition, Clarendon Press, Oxford
(2002).

\bibitem{Ma}
S. K. Ma, {\it Modern Theory of Critical Phenomena}, Chpt. IX, W. A. Benjamin Inc., Reading, Mass. (1976);

\bibitem{heat}
This type of transition was first observed in the fluctuations of the heat exchanged
by a ferromagnet quenched below the critical point in Ref.~\cite{US} and in the 
fluctuations of composite operators whose average are correlation and response
functions in Ref.~\cite{FL}. 


\bibitem{driveBK}
In the usual treatment of the spherical model~\cite{BK}, one fixes $s=1$ and the transition is driven
by $T$. Conversely, here $T_I$ is fixed and the transition is
driven by $s$.

\bibitem{Weitz}
J. Klaers, J. Schmitt, F. Vewinger and M.Weitz, Nature {\bf 468}, 545 (2010)

\bibitem{Godreche}
C. Godr\`eche and J. M. Luck, J. Phys. A: Math. Gen. {\bf 33}, 9141 (2000)
F. Corberi, E. Lippiello and M. Zannetti, Phys. Rev. E, {\bf 65} 046136 (2002);
A. Annibale and P. Sollich, J. Phys. A: Math. Gen. {\bf 39}, 2853 (2006)









\end{thebibliography}
\end{document}